\documentclass[twocolumn,twoside,slac_two]{revtex4}
\usepackage{graphicx}
\usepackage{fancyhdr}
\def\xsrcnos{4U~0142+61}
\def\xsrc{4U~0142+61~}
\def\xsrcnos{4U~0142+61}

\def\rxs{1RXS~J170849.0-400910~}
\def\esrcs{1E~1841-045~}

\def\iglnos{INTEGRAL}

\begin{document}

\title{High-Energy Gamma-ray Emission Properties of an Anomalous X-ray Pulsar, \xsrcnos}

%

\author{S. \c{S}a\c{s}maz Mu\c{s}, E. G\"o\u{g}\"u\c{s}}
\affiliation{Sabanc\i~University, FENS, Orhanl\i - Tuzla, 34956, Istanbul, Turkey}
\begin{abstract}
Anomalous X-ray Pulsars (AXPs) are bright X-ray sources. Few AXPs emit highly pulsed emission in hard X-rays. Using data collected with the Large Area Telescope on board Fermi Gamma-ray Space Telescope, we explored high-energy gamma-ray emission from the brightest AXP, \xsrcnos. We do not detect any significant emission from the source. Here, we present the upper limits to the persistent and pulsed emission of \xsrc in the high-energy gamma-ray domain.  
\end{abstract}

\maketitle

\thispagestyle{fancy}


\section{introduction}

Anomalous X-ray Pulsars (AXPs) are bright X-ray emitters; their luminosities (below 10 keV) range from $10^{33}$ to $10^{36}$ erg s$^{-1}$ that exceed their spin-down luminosity. Their spin periods are clustered between 2$-$12 s and spin-down rates are relatively large, i.e., between $10^{-10}$$-$$10^{-13}$ s s$^{-1}$ (see \citep{mereghetti08} for a review). Their observational properties are explained either with a fallback disk \citep{CHN00, Alpar01} or by the decay of their strong magnetic field (magnetar model) \citep{td96}. 

Until the discovery of hard X-ray emission from AXPs \esrcs \citep{kuiper04}, \xsrc \citep{dH04} and \rxs \citep{revn04} they were known to be emitting soft X-rays only ($<$ 10 keV). The origin of hard X-ray emission is still not well understood. Emission from corona \citep{bt07}, breakdown of fast-modes via quantum electrodynamics effects \citep{hh05a, hh05b} and resonant Compton upscattering of soft photons \citep{bh07} are proposed to explain the hard-X ray emission.

Construction of spectral behaviors of these sources in a broad energy band is important for understanding the origin of the hard X-ray emission. Here we present the upper limits to the persistent and pulsed high-energy gamma-ray emission from \xsrc at the GeV range. We estimated an upper limit to the spectral break energy by extending the $\nu$F$_{\nu}$ spectrum of the source to GeV range. 

\begin{figure*}[t]
\centering
\includegraphics[width=135mm]{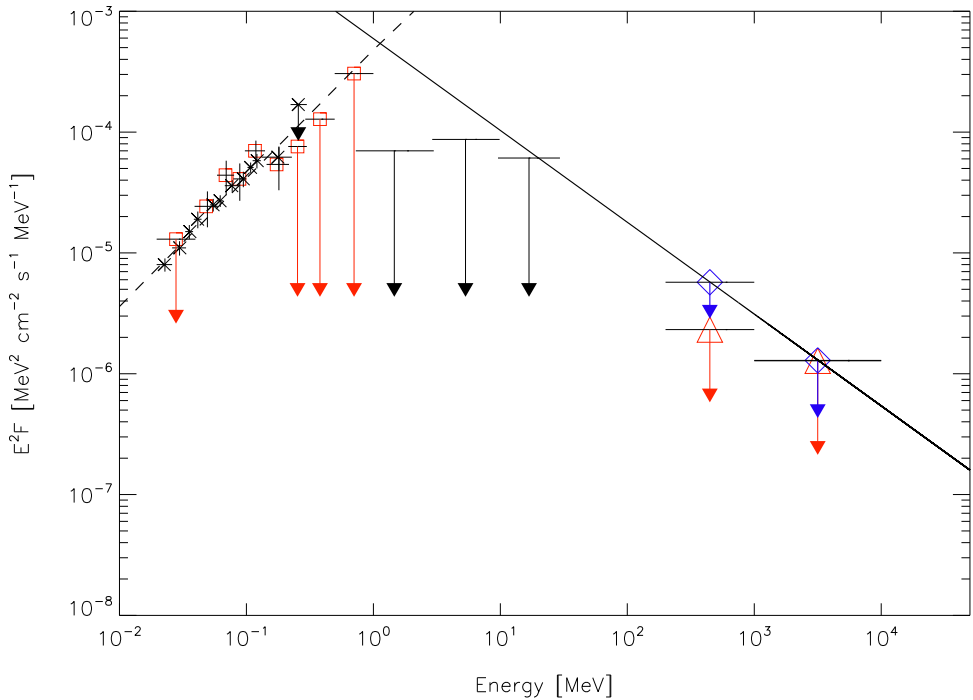}
\caption{A wide band $\nu$F$_{\nu}$ spectrum of 4U 0142+61: \textit{\iglnos}/ISGRI (20-300 keV) in black (stars), \textit{\iglnos}/SPI (20-1000 keV) in red (open squares) and \textit{CGRO}/COMPTEL (0.75-30 MeV) 2$\sigma$ upper limits in black 
(data obtained from \citep{dH08}). Blue diamonds are the \textit{Fermi}/LAT upper limits in the 0.2$-$1.0 GeV and 1.0$-$10.0 GeV obtained using $2^{\circ}$ extraction region. Red triangles are upper limits for the $15^{\circ}$ extraction region. Dashed line is the best fit power-law model to the ISGRI data points \citep{dH08}. Solid line shows the power-law upper limit trend of the $2^{\circ}$ \textit{Fermi}/LAT region. Figure is from \citep{ssm10}}
\label{fig:nufnu}
\end{figure*}

\section{Observations and Data Analysis}

The LAT observations between 2008 August 4 to 2010 April 29 with an exposure time of $\sim$31.7 Ms were used to investigate the persistent emission from \xsrcnos. We obtained and analyzed the data from $15^{\circ}$ radius 
around the source and also, a $2^{\circ}$ radius region was selected and analyzed in order to avoid contamination from nearby bright sources. Spectral fits and flux calculations were done with the Python version of gtlike, pyLikelihood, for the 0.2$-$1.0 and 1.0$-$10.0 GeV energy bands. See \citep{ssm10} for the details of LAT data calibration.

To perform timing analysis, all event photon arrival times were extracted from a $2^{\circ}$ region and converted by gtbary to the arrival times at the solar system barycenter. We used 2$-$10 keV \textit{RXTE} observations and 
found the spin ephemeris of the source between 2008 August 4 and 2010 April 30 which also covers the extracted LAT observations.

To search for pulsed high-energy gamma-ray emission from \xsrcnos, first we obtained the spin ephemeris of the source using contemporaneous RXTE/PCA observations in the 2$-$10 keV range with a total exposure of 196 ks. A Fourier based epoch folding technique was applied to the data to obtain the spin ephemeris. We generated the pulse profiles of the source using three consecutive PCA observations around the epoch (MJD 54713.5) which are grouped such as they are separated at least 0.2 days from each other. We determined the phase shift of each pointing with respect to the template by cross correlating the pulse profiles of each group of pointings with
the template profile and fitted the phase shifts with a polynomial. In Table 1 we present the best fit spin ephemeris parameters of \xsrcnos. We used the precise PCA spin ephemeris that we obtained to search for pulsed high-energy gamma-ray emission from \xsrcnos. We generated the LAT pulse profiles in the 0.2$-$1.0 GeV and 1.0$-$10.0 GeV energy ranges and found that both LAT profiles are consistent with random fluctuations with respect to its mean. 

\begin{table}[h]
\begin{center}
\caption{Spin ephemeris of \xsrc as determined using RXTE/PCA observations.}
\vspace{0.5cm}
\begin{tabular}{lc}\hline\hline
Parameter                   & Value   \\ \hline
Range (MJD)  &  54682.6 $-$ 55315.1 \\
Epoch (MJD)  	  & 54713.5  \\
$\nu$ (Hz)        & 0.1150900026(9) \\
$\dot{\nu}$ ($10^{-14}$ Hz s$^{-1}$)  &  $-$2.745(8) \\
$\ddot{\nu}$ ($10^{-23}$ Hz s$^{-2}$) &   3.6(3) \\ \hline
\end{tabular}
\label{tab:poly_fit}
\end{center}
\vspace{-0.5cm}
\end{table}

\section{Results}
\begin{figure*}
\hspace{0.3cm}
\includegraphics[width=53mm]{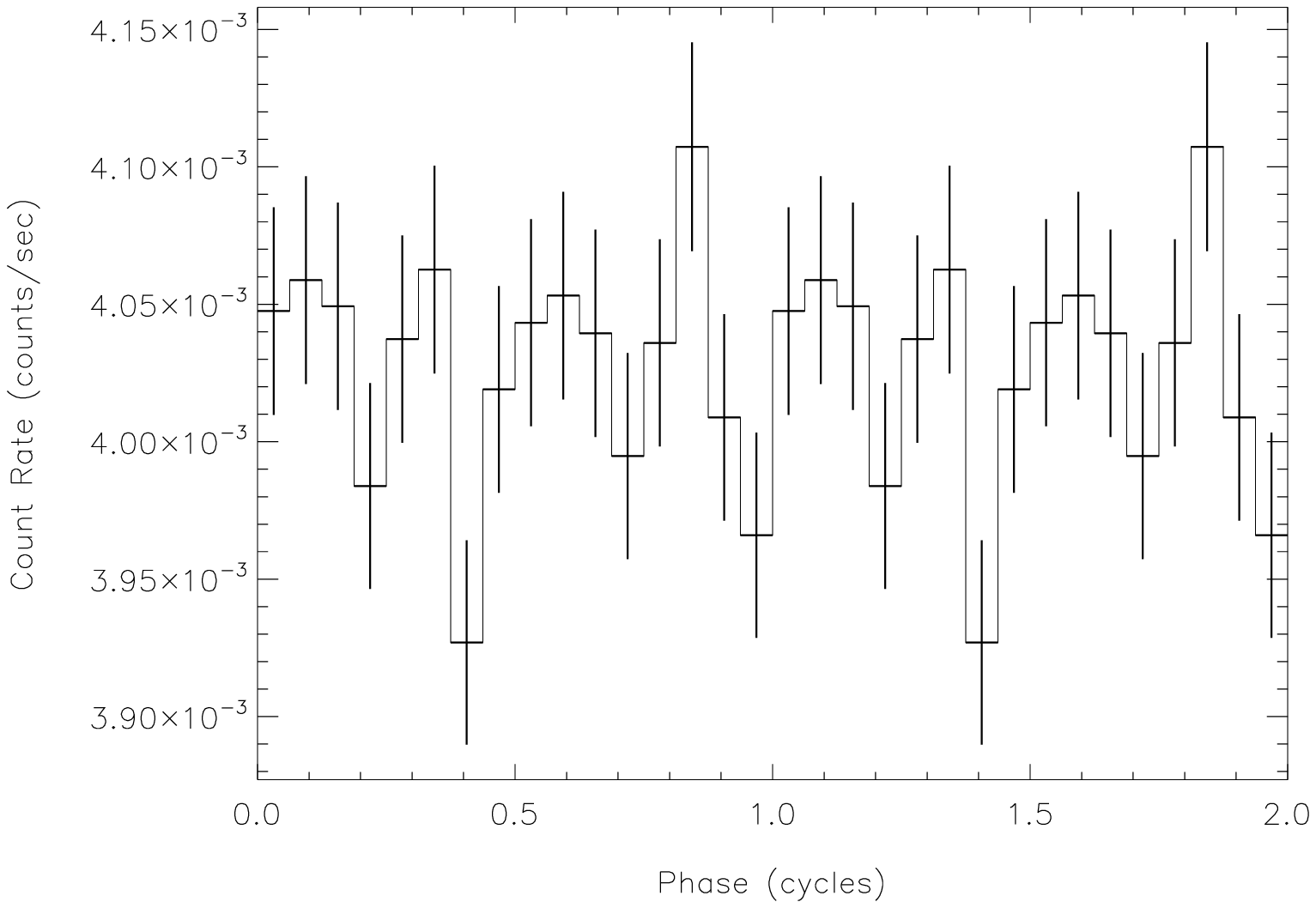}%
\hspace{0.2cm}%
\includegraphics[width=53mm]{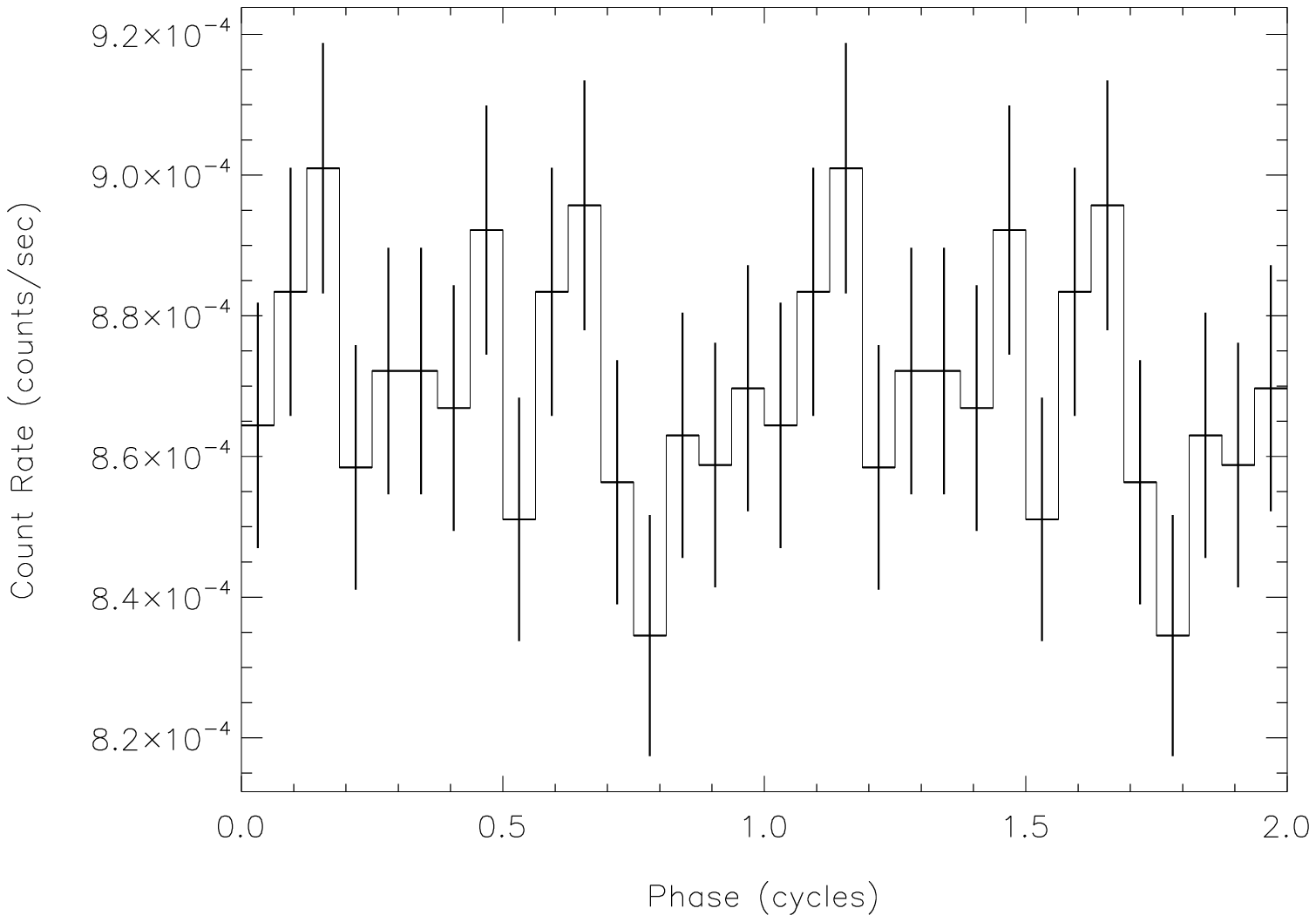}%
\hspace{0.2cm}%
\includegraphics[width=53mm]{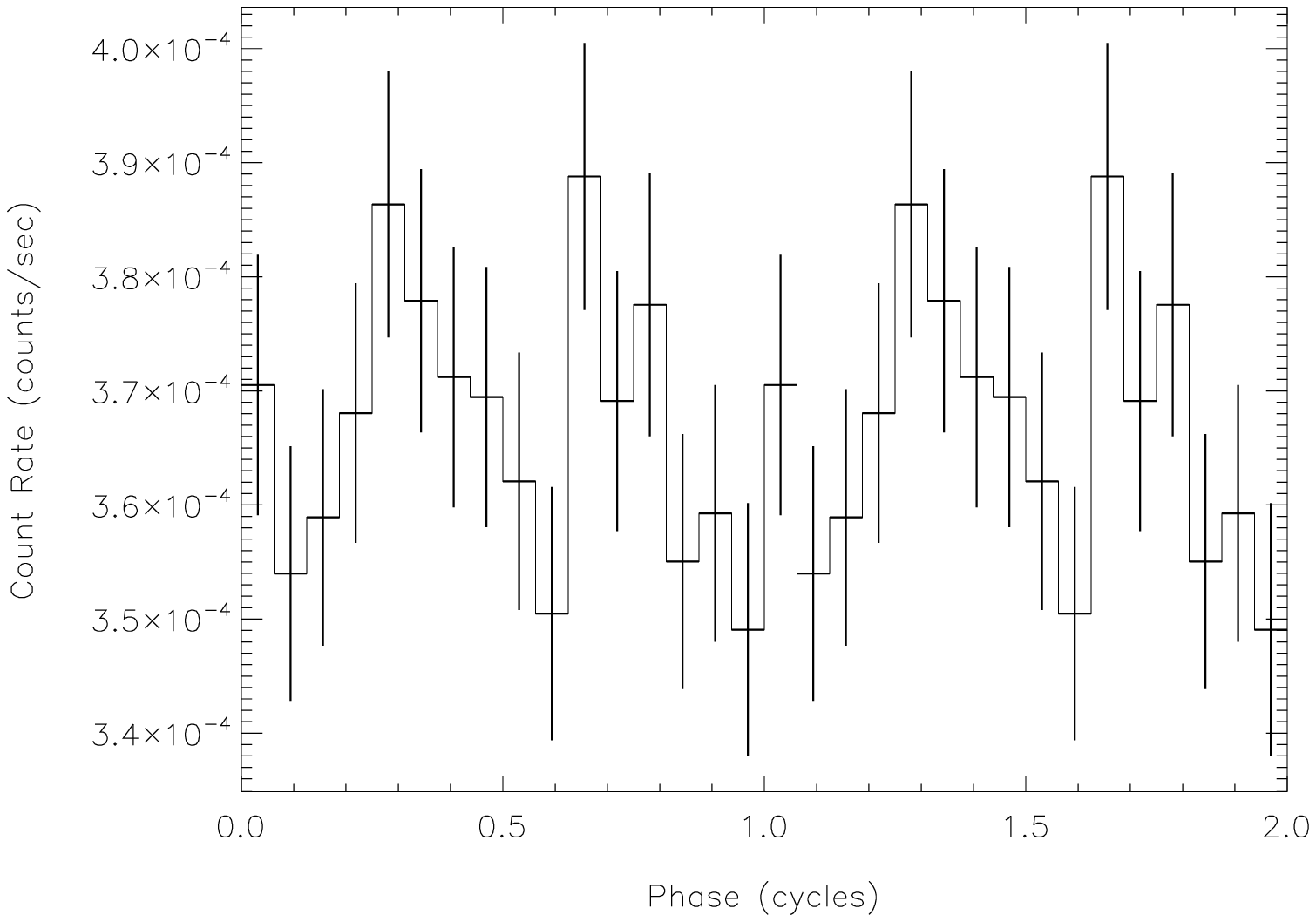} %
\caption{Pulse Profiles of \xsrc by folding the Fermi/LAT data with the precise spin ephemeris obtained with contemporaneous RXTE/PCA observations. \textit{Left}: in the 30 $-$ 200 MeV band, \textit{middle}: in the 200 MeV $-$ 1 GeV band, \textit{Right}: in the 1 $-$ 10 GeV band.}
\label{fig:profiles}
\end{figure*}

After processing the data as explained in \S 2, we fitted a power-law to the data obtained from $15^{\circ}$ radius region with an index of 2.5. The fit results a test statistics (TS) value of $\sim$0.23 which implies a detection significance less than 1$\sigma$. We calculated 3$\sigma$ flux upper limits as 2.32 $\times$ 10$^{-6}$ MeV cm$^{-2}$ s$^{-1}$ in the 0.2$-$1.0 GeV band and 1.28 $\times$ 10$^{-6}$ MeV cm$^{-2}$~s$^{-1}$ in the 1.0$-$10.0 GeV band. For the $2^{\circ}$ radius region power-law fit with an index of 3 resulted in a TS value of $\sim$3 which implies a detection significance less than 2$\sigma$. The 3$\sigma$ flux upper limits are  5.72 $\times$ 10$^{-6}$ MeV cm$^{-2}$ s$^{-1}$ and 1.29 $\times$ 10$^{-6}$ MeV cm$^{-2}$ s$^{-1}$ for 0.2$-$1.0 GeV and 1.0$-$10.0 GeV, respectively.\\
\\
We searched for pulsed high-energy gamma-ray emission from \xsrc by folding the Fermi/LAT data with the precise spin ephemeris obtained using contemporaneous RXTE/PCA observations. Our search for emission yields random fluctuations with respect to its mean (See Figure \ref{fig:profiles}).
The 3$\sigma$ upper limits to the RMS pulsed amplitude are 1.5\% and 2.3\% in the 0.2$-$1.0 GeV and  
1.0$-$10.0 GeV band, respectively. A search in the lower energy part of the LAT passband (30$-$200 MeV) also yields in no evidence of pulsed emission; the 3$\sigma$ RMS pulsed amplitude upper limit is 1.6\%. 

\section{Discussion}
We used the LAT persistent emission upper limits to extend the $\nu$F$_{\nu}$ spectrum of \xsrc presented in \citep{dH08} to the 10 GeV. We could place an upper limit to the very high-energy gamma-ray emission trend (i.e., the LAT results obtained using the $2^{\circ}$ region) which is a power law with an index of $-$0.76 (solid line in Figure \ref{fig:nufnu}). \textit{\iglnos}/ISGRI data is fitted with a power law index of 0.93 $\pm$ 0.06 by \citep{dH08} as shown with dashed line in Figure \ref{fig:nufnu}. The intersection of two curves suggests an upper limit of $\sim$1.1 MeV to the spectral break energy.
 
The models presented in \citep{bt07} and \citep{hh05a, hh05b} estimate a spectral break energy of about 1 MeV which is consistent with the break energy that we estimated. If there is a significant excess emission in optical band as in the \xsrc \citep{hulleman00}, quantum electrodynamics model \citep{hh05a, hh05b} predicts an increase in the 10 $-$ 200 MeV range. Excess optical emission can be due to the star itself \citep{km02, dhillion05} or the disk around the star \citep{wang06, es06}. If the source of the excess optical emission is the disk, quantum electrodynamics model is consistent with our results since the star itself would not be excessive in optical band.
Outer gap model of AXPs within the magnetar interpretation \citep{CZ01} also predicts high-energy gamma-ray emission from the source. Recently, \citep{TSX10} and \citep{TSX11} suggested that non-detection is inconsistent with the outer gap model and fallback interpretation cannot be eliminated.

\bigskip 
\begin{acknowledgments}
S.\c{S}.M. and E.G. acknowledge EU FP6 Transfer of Knowledge Project, 
Astrophysics of Neutron Stars (MTKD-CT-2006-042722).
\end{acknowledgments}

\bigskip 

\end{document}